\documentclass[aps,prl,superscriptaddress,reprint]{revtex4-1}
\usepackage{amsmath} \usepackage{amsfonts}
\usepackage{amssymb} 
\usepackage{array}
\usepackage{graphicx} 
\usepackage{verbatim}

\begin{document}

\preprint{APS/123-QED}
\title{Magnetotransport patterns of collective localization near $\nu=1$ \\
in a high-mobility two-dimensional electron gas}


\author{S.A. Myers}
\affiliation{Department of Physics and Astronomy, Purdue University, West Lafayette, IN 47907}
\author{Haoyun Huang}
\affiliation{Department of Physics and Astronomy, Purdue University, West Lafayette, IN 47907}
\author{L.N. Pfeiffer}
\affiliation{Department of Electrical Engineering, Princeton University, Princeton, NJ 08544}
\author{K.W. West}
\affiliation{Department of Electrical Engineering, Princeton University, Princeton, NJ 08544}
\author{G.A. Cs\'athy}
\affiliation{Department of Physics and Astronomy, Purdue University, West Lafayette, IN 47907}

\date{\today}
             
\begin{abstract}

We report complex magnetotransport patterns of the $\nu=1$ integer quantum Hall state in a GaAs/AlGaAs sample from the newest generation with a record high electron mobility. The reentrant integer quantum Hall effect
in the flanks of the $\nu=1$ plateau indicates the formation of the integer quantum Hall Wigner solid, a collective
insulator. Moreover, at a fixed filling factor, the longitudinal resistance versus temperature in the region of the 
integer quantum Hall Wigner solid exhibits a sharp peak. Such sharp peaks in the longitudinal resistance versus temperature
so far were only detected for bubble phases forming in high Landau levels
but were absent in the region of the Anderson insulator. We suggest that in samples of sufficiently low disorder
sharp peaks in the longitudinal resistance versus temperature traces are universal transport signatures 
of all isotropic electron solids that form in the flanks of integer quantum Hall plateaus.
We discuss possible origins of these sharp resistance peaks and
we draw a stability diagram for the insulating phases in the $\nu$-$T$ phase space.

\end{abstract}

\maketitle

\section{Introduction: single electron and collective localization on an integer quantum Hall plateau}

Electron localization plays an important role in topological materials. 
This is because transport supported by the boundary states of these materials is protected  when
electrons in the bulk are localized. These ideas \cite{laughlin,aoki,halperin} were introduced in order to explain the  
plateaus of integer quantum Hall states (IQHS)  that form in 
the two-dimensional electron gas (2DEG) \cite{klitz}, and then later extended to other topological materials.

Over the years, the concept of electron localization has evolved.  It was realized that 
in order to understand localization along a plateau of an IQHS, both single particle and
collective localization had to be invoked. According to current understanding,
an Anderson insulator (AI) forms near the center of an integer quantum Hall plateau
at low quasiparticle densities.
However, as the quasiparticle density is increased past the range of the AI, electrons 
in 2DEGs of sufficiently low disorder
reorganize themselves in collective insulators called charge density waves \cite{fogler,moessner,fogler96,fogler97,haldane}.
These charge density waves are pinned by the disorder, hence their insulating behavior.
One example of such a collective insulator is the
Wigner solid forming in the flanks of an integer quantum Hall plateau,
also referred to as the integer quantum Hall Wigner solid (IQHWS) \cite{chen-int,lewis-1,lewis-2};
the IQHWS is related to the Wigner solid forming at the largest magnetic fields \cite{ws1,ws2,ws3}. 
In high Landau levels, a further increase of the quasiparticle density
leads to the formation of the electronic bubble phases (BPs) and of stripe phases
\cite{fogler,moessner,fogler96,fogler97,haldane,du,lilly,cooper,eisen02,zudov,kevin1}. 
These collective insulators were discovered in 2DEGs confined to GaAs/AlGaAs interfaces
\cite{chen-int,lewis-1,lewis-2,ws1,ws2,ws3,du,lilly,cooper,eisen02,zudov,kevin1}, 
but recently BPs \cite{dean} and IQHWSs \cite{young}  were also observed in high quality graphene.

The isotropic insulating phases near an integer plateau, i.e. the AI, IQHWS, and BPs, all exhibit the same transport behavior:
they have a vanishing magnetoresistance $R_{xx}=0$ and a Hall resistance quantized
to the value of a nearby integer plateau, $R_{xy}=h/i e^2$, where $i$ is an integer. 
Nonetheless, BPs are separated from other localized states that form in the middle of an integer plateau
by a conspicuous deviation of magnetotransport from quantization. 
Such a behavior is commonly referred to as reentrance of the integer quantum Hall effect\cite{du,cooper}.
In contrast, a similar reentrant transport is typically absent between the IQHWS and the AI.
Therefore in order to distinguish the IQHWS from the AI, techniques other than transport
were developed. Examples of such techniques are absorption in the microwave frequency 
domain \cite{chen-int,lewis-1,lewis-2}, compressibility \cite{smet-int},
nuclear magnetic resonance \cite{tiemann-int}, surface acoustic wave propagation \cite{suslov-int},
and tunneling measurements \cite{ashoori-int}.  The continued absence of any features 
in magnetoresistance separating the IQHWS and the AI brought into question 
whether or not dc transport could successfully differentiate the IQHWS from the AI.

Transport features separating the IQHWS from the middle part of the integer quantum Hall
plateau akin to reentrance of the integer quantum Hall effect
were reported in low disorder GaAs samples \cite{liu-1} and more recently in graphene \cite{young}.
However, there is limited data available in this region.
Here we present a collection of stunning magnetotransport patterns from the newest generation of ultralow disorder 
GaAs samples\cite{chung}. Reentrant transport in the flanks of the $\nu=1$ IQHS is associated with the
IQHWS. We further investigate the IQHWS by analyzing the longitudinal resistance versus temperature 
traces at fixed filling factors, or $R_{xx}(T)|_{\nu=\text{fixed}}$. We find that 
$R_{xx}(T)|_{\nu=\text{fixed}}$ traces exhibit a sharp peak in the IQHWS region, but such sharp peaks
are absent for the AI. The temperature of this sharp peak is interpreted as the onset of the IQHWS.
In higher Landau levels, sharp peaks that resemble those reported in the present work
have been observed for all multi-electron BPs \cite{deng-1,deng-2,chick,kevin2}. 
We thus surmise that sharp peaks in  $R_{xx}(T)|_{\nu=\text{fixed}}$ are likely a universal property
of the isotropic electron solids. In the following we discuss possible origins of such sharp resistance peaks and
we draw a stability diagram for the AI and the IQHWS forming near $\nu=1$ in the $\nu$-$T$ phase space.

\section{Sample details and experimental techniques}

Recent efforts of source-material purification and innovation in the GaAs molecular beam epitaxy technique
produced 2DEGs with record high mobilities\cite{chung}. Our measurements were performed on a sample
that belongs to this newest generation, with an electron density
$n = 7.5\times10^{10} \text{ cm}^{-2}$ and a low temperature mobility of $\mu = 24 \times 10^6 \text{ cm}^2/\text{V s}$. At this electron density the mobility exceeds that of samples from an earlier generation 
by more than a factor of two \cite{chung}. 
The width of the confining quantum well is $58.5$~nm.
Our sample was cleaved to $4\times4 \text{ mm}^2$ size with eight indium contacts in the van der Pauw geometry,
and it was mounted in a He-3 immersion cell\cite{setup} in order to stabilize the electron temperature. 
The sample state was prepared by a brief illumination with a red light emitting diode.
Simultaneous longitudinal resistance $R_{xx}$ and Hall resistance $R_{xy}$ measurements were performed 
using two lock-in amplifiers with a current excitation of $3$~nA and a frequency of $13.3$~Hz.

\begin{figure}[t]
  \includegraphics[width=1\columnwidth]{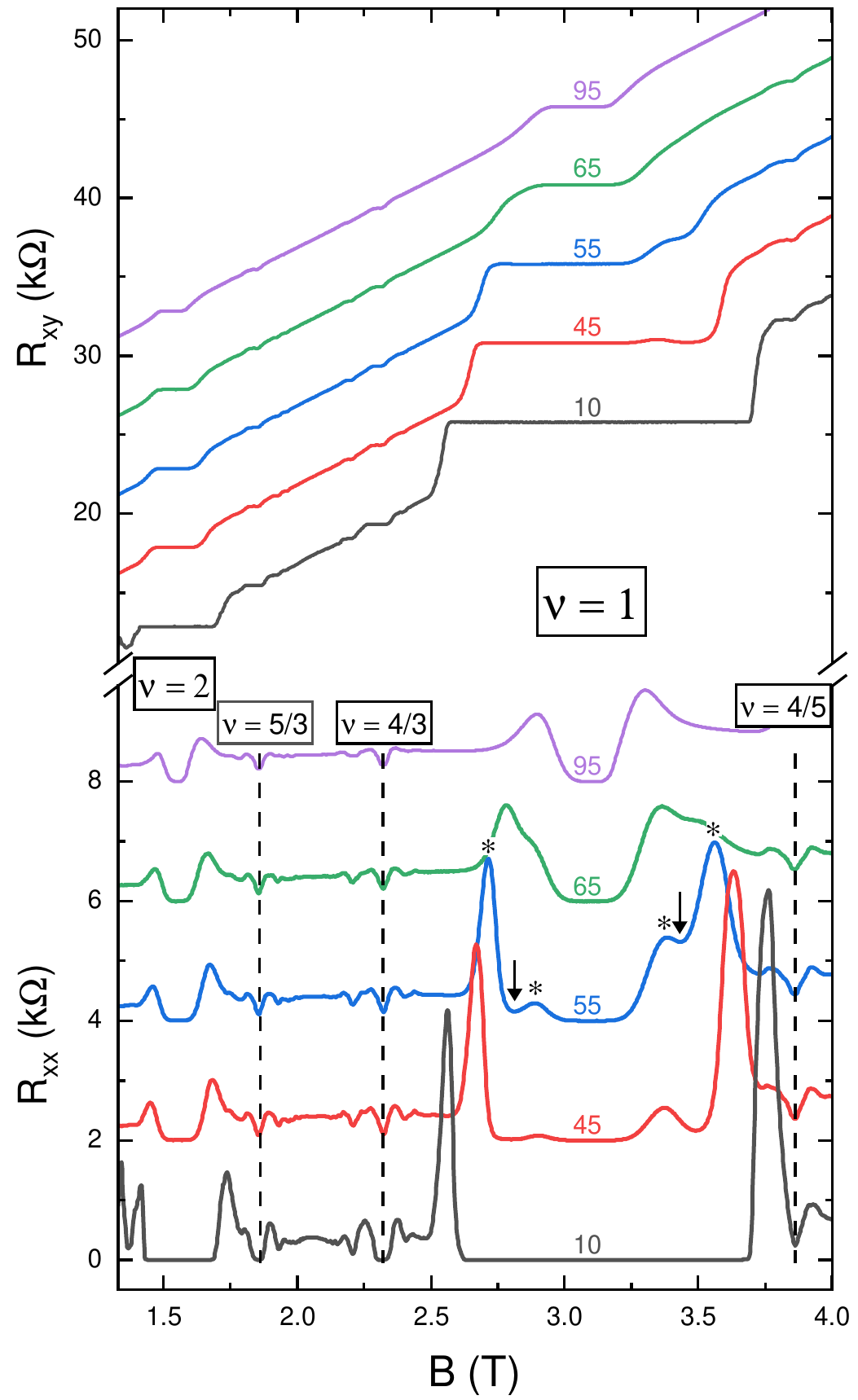} 
  \caption{The evolution of the longitudinal magnetoresistance $R_{xx}$ and Hall resistance $R_{xy}$ plotted against  the magnetic field $B$ at several temperatures. 
Labels in boxes mark integer and fractional quantum Hall states, whereas numbers indicate
temperatures in mK.  $R_{xx}$ traces other than the one at $T=10$~mK are offset by 2~k$\Omega$. Similarly,
$R_{xy}$ traces other than the one at $T=10$~mK are offset by 5~k$\Omega$.
For the $R_{xx}$ trace at $T=55$~mK, arrows mark reentrant integer quantum Hall states associated with the IQHWS,
while the $\ast$ symbols mark consecutive local maxima in $R_{xx}$ between which the IQHWSs form. 
\label{fig:1} }
\end{figure}

\begin{figure}[t]
  \includegraphics[width=\columnwidth]{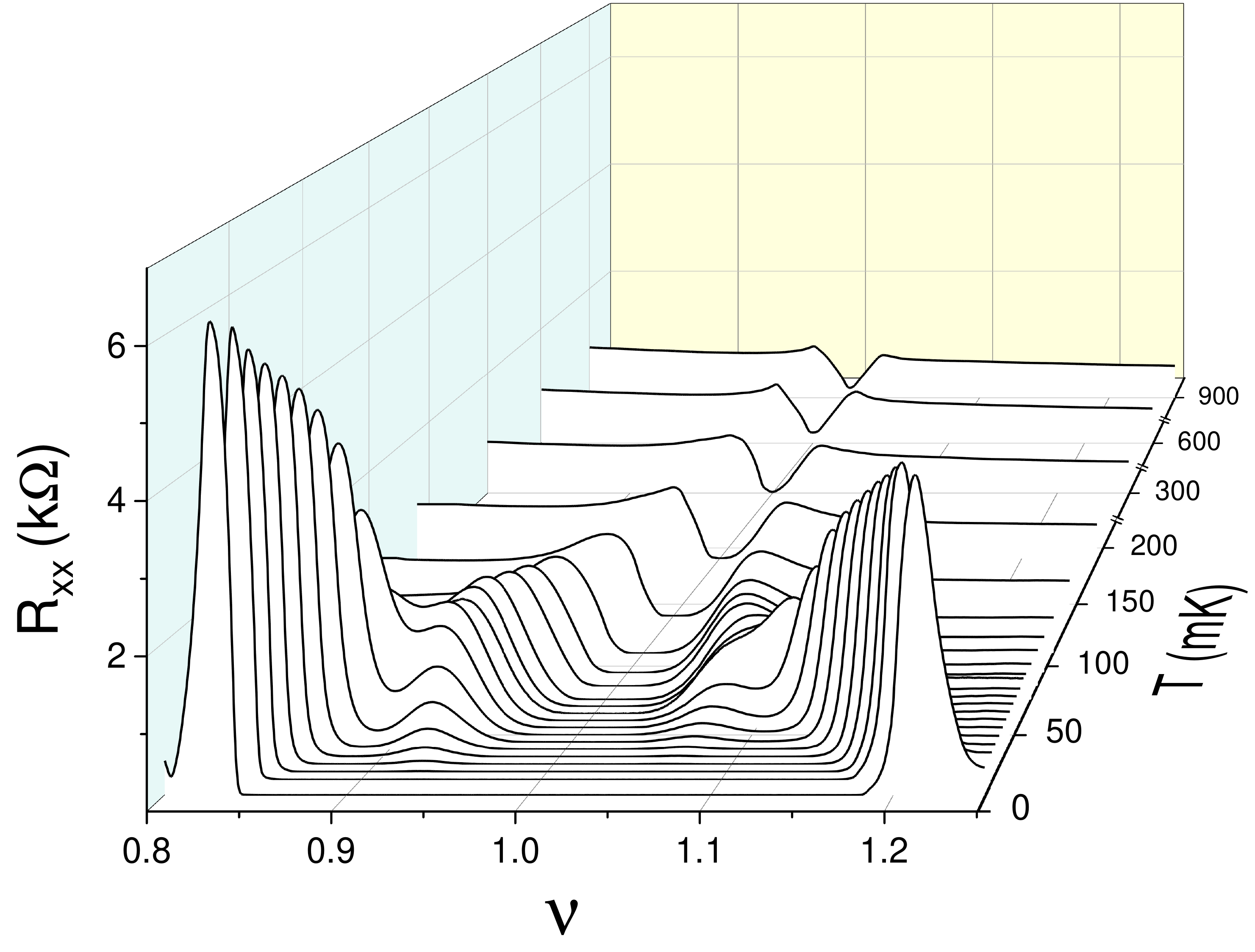}
  \caption{The evolution of the longitudinal magnetoresistance $R_{xx}$ at $\nu=1$
   in the $\nu$-$T$ parameter space for temperatures ranging from 10 to 900 mK. 
   Line breaks were added between high temperature traces.  
\label{fig:2} }
\end{figure}

\section{Magnetotransport near the $\nu=1$ integer quantum Hall plateau}

The longitudinal magnetoresistance $R_{xx}$ and the Hall resistance $R_{xy}$ 
are measured at several temperatures. Representative traces are shown in Fig.\ref{fig:1}. In this figure
we marked the IQHSs at $\nu=1$ and $2$ as well as the fractional quantum Hall states observed
at $\nu=5/3$, $4/3$, and $4/5$. 
At $T=95$~mK, the measured magnetoresistance $R_{xx}(B)|_{T=\text{fixed}}$ close to $\nu=1$
is typical for that of an integer quantum Hall state:
$R_{xx}$ vanishes over a range of magnetic fields of about $0.2$~T and $R_{xy}=h/e^2$ in the same range of fields.  
However, as the temperature is lowered, 
a more complex structure develops. For example, at $T=55$~mK there is a local minimum developing in $R_{xx}$
near $B=3.5$~T. This local minimum is marked by an arrow in Fig.\ref{fig:1} and
it is located between two local maxima marked by the $\ast$ symbols.
A further lowering of the temperature to $T=45$~mK results in a deeper resistance minimum in $R_{xx}$ 
and in a fully quantized $R_{xy}$ near $B=3.5$~T.
The sequence of vanishing $R_{xx}$ in the center of the integer plateau, a local maximum in $R_{xx}$
near $B=3.3$~T, and the vanishing $R_{xx}$ near $B=3.5$~T signals a reentrance of the integer quantum Hall 
plateau. Following arguments put forth earlier\cite{du,cooper,liu-1}, 
such a reentrant behavior is not expected for the AI and hence
the region of the local minimum in $R_{xx}$ near $B=3.5$~T is associated with a collective insulator.
We observe a similar collective insulator on the other flank of the $\nu=1$ integer quantum Hall plateau near $B=2.75$~T.

For insight on the nature of the collective insulators in the flanks of the $\nu=1$ IQHS, we examine the
Landau level filling factors at which they develop. 
In Fig.\ref{fig:2} we show a series of $R_{xx}$ traces against $\nu$ over an extended range of temperatures. 
The collective insulators near $B=2.75$~T and $B=3.5$~T are seen in Fig.\ref{fig:2}
near the filling factors $\nu=1.1$ and $\nu=0.9$.
A more careful analysis shows that the collective insulators at the lowest temperatures develop 
in the range of filling factors $0.83 \lesssim \nu \lesssim 0.93$ and $1.06 \lesssim \nu \lesssim 1.21$.
IQHWSs are known collective insulators forming in the above ranges of filling factors.
Indeed, the above ranges of filling factors overlap extremely well with those
of IQHWSs observed in microwave absorption \cite{chen-int} and in 
tunneling measurements \cite{ashoori-int}, and they are reasonably close to
those from compressibility \cite{smet-int} and transport measurements \cite{liu-1}.
Furthermore, according to Hartree-Fock calculations \cite{fogler,fogler96,moessner,fogler97},
single-electron BPs, i.e. Wigner solids, are the only collective insulators that may form
in the lowest Landau levels at the above ranges of filling factors.
Therefore, based on transport signatures, results of prior experiments \cite{chen-int, ashoori-int, smet-int,liu-1},
and based on consistency with theory, we interpret the collective ground states 
 near the filling factors $\nu=1.1$ and $\nu=0.9$ as IQHWSs.
Furthermore, we associate the part of the integer quantum Hall plateau between the two reentrant states with AI.

\begin{figure}[t]
  \includegraphics[width=\columnwidth]{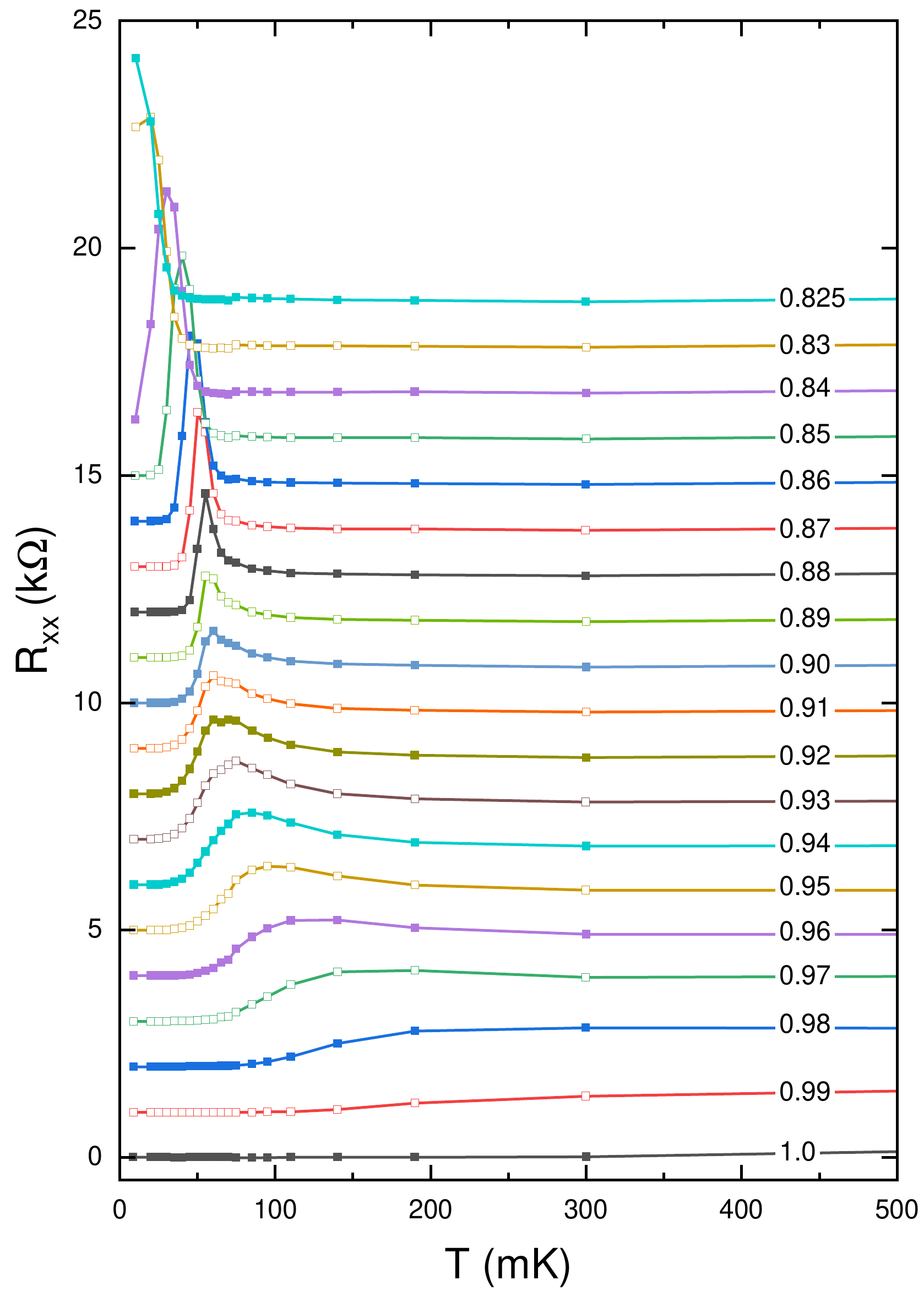}
  \caption{Waterfall plot of $R_{xx}(T)|_{\nu=\text{fixed}}$ resistance profiles
  for $\nu=1.0$ to 0.825. With the exception of the $\nu=1$ trace, traces are vertically offset  by 1~k$\Omega$. 
  The sharp resistance peaks that develop at $\nu \leq 0.90$ are associates with the IQHWS.
\label{fig:3} }
\end{figure}

In the following, we analyze the temperature evolution of the magnetoresistance 
of the $\nu=1$ integer quantum Hall plateau.
We extract cuts along constant $\nu$ in the complex landscape of $R_{xx}(\nu,T)$ shown in Fig.\ref{fig:2}.
The resulting $R_{xx}(T)|_{\nu=\text{fixed}}$ resistance profiles obtained for various filling
factors are shown in Fig.\ref{fig:3}. Here we focus on the behavior of the $R_{xx}(T)|_{\nu=\text{fixed}}$
resistance profiles for $\nu \leq 1$; however, we note that a qualitatively similar trend is also observed for $\nu \geq 1$
(not shown in Fig.\ref{fig:3}). For most traces in Fig.\ref{fig:3} $R_{xx}=0$ at $T=10$~mK, and thus the
corresponding filling factors are associated with the plateau of the $\nu=1$ IQHS.
The most striking feature of the $R_{xx}(T)|_{\nu=\text{fixed}}$ resistance profiles is the sharp peak
present at filling factors $\nu \leq 0.90$. 
At other filling factors, such as the ones in the $0.97 \leq \nu \leq 1$ range, 
the resistance profiles $R_{xx}(T)|_{\nu=\text{fixed}}$ display a gradual and monotonic decrease
as $T$ is lowered. 
Yet at other filling factors, such as at $\nu \approx 0.96$, in $R_{xx}(T)|_{\nu=\text{fixed}}$
there is a small and relatively broad local maximum;
the peak resistance at $\nu \approx 0.96$ is barely above the flat background resistance value measured
at higher temperatures. 
Since sharp peaks in the $R_{xx}(T)|_{\nu=\text{fixed}}$ resistance profile occur in the range
of filling factors of the IQHWS, we surmise that they are a fundamental signature in the dc transport of this state.

Sharp peaks in $R_{xx}(T)|_{\nu=\text{fixed}}$ that resemble the ones shown in Fig.\ref{fig:3} for the IQHWS
are also present for the BPs residing in the second and higher Landau levels \cite{deng-1,deng-2,chick,kevin2},
but are absent for the AI.
We therefore suggest that a sharp peak in $R_{xx}(T)|_{\nu=\text{fixed}}$ is a shared property of all
isotropic collective insulators forming in the flanks of integer quantum Hall plateaus, i.e.
the IQHWSs and BPs at least in a restricted parameter space. 
We note that since the IQHWS is identical to the one-electron BP \cite{fogler,moessner,fogler96,fogler97}, 
it is reasonable that these phases, i.e. the multi-electron and single-electron BPs,
share similar transport properties. The temperature of the sharp peak in
$R_{xx}(T)|_{\nu=\text{fixed}}$ can then be interpreted as the onset
temperature $T_{\text{onset}}$ of these collective insulators \cite{deng-1,deng-2,chick,kevin2}.
The largest onset temperature of IQHWS near $\nu=0.90$ in our sample is $T_{\text{onset}}=60$~mK,
whereas near $\nu=1.1$ $T_{\text{onset}}=65$~mK.

\begin{figure}[t]
  \includegraphics[width=\columnwidth]{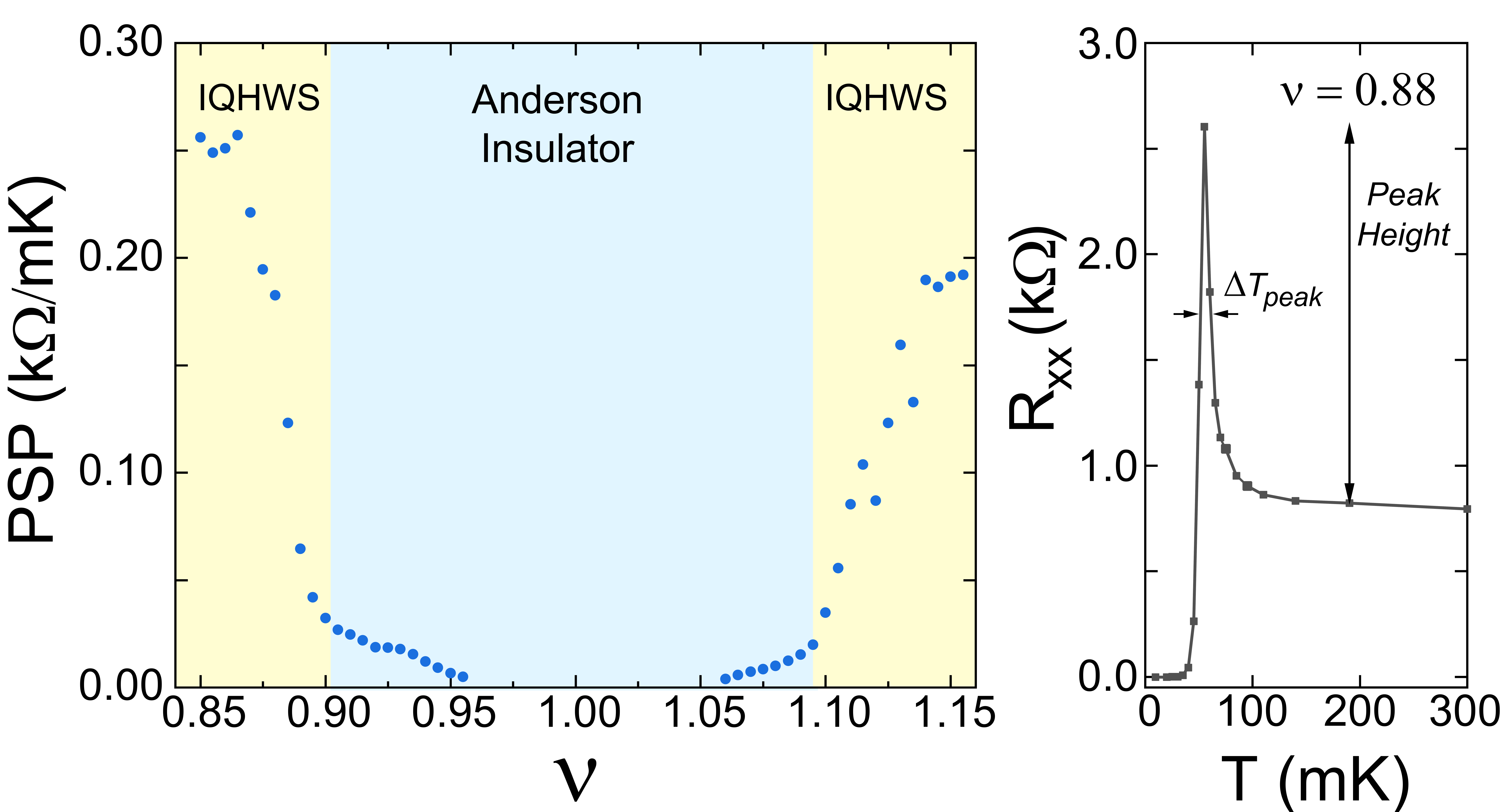}
  \caption{Plot of the peak sharpness parameter (PSP) plotted as a function of filling factor (left). $R_{xx}(T)$ at $\nu=0.88$ (right) provides reference for the quantities used in calculating the PSP.\label{fig:4}}
\end{figure}

To quantify the trends in the sharp resistive peaks in $R_{xx}(T)|_{\nu=\text{fixed}}$ 
and shown in Fig.\ref{fig:3}, we 
define the peak sharpness parameter (PSP) as the ratio of the $R_{xx}$ peak height 
relative to $R_{xx}(T=190 \text{mK})$, a measure of the flat resistive background at high temperatures,
and the width $\Delta T_{\text{peak}}$ at half height, as measured in mK.
Such a parameter becomes large when the peak becomes prominent, i.e.  
when the peak height increases and when its width becomes narrow. The PSP
is thus useful for quantifying the sharpening of resistance peaks seen in Fig.\ref{fig:3}.
PSP values extracted this way are plotted as a function of the filling factor in Fig.\ref{fig:4}. 
In the $0.90 < \nu < 0.955$ range, the PSP changes little with $\nu$.  In this range of filling facors 
$R_{xx}(T)|_{\nu=\text{fixed}}$ profiles have the small, broad peak present.
For $\nu < 0.90$ where the $R_{xx}(T)|_{\nu=\text{fixed}}$ profiles exhibit a sharp peak,
there is a precipitous increase of the PSP parameter. The sudden increase
of the PSP parameter at $\nu=0.90$ thus signals the onset of the IQHWS. 
At larger filling factors we find a similar onset of the IQHWS near $\nu=1.095$.
Each of these two filling factors can be thought of as the demarcation point between the IQHWS and the AI
on the filling factor axis. 

In our interpretation, at $T \ll T_{\text{onset}}$ the IQHWS is stable and it is fully localized,
at $T \gg T_{\text{onset}}$ the ground state is an electron liquid, and at $T \approx T_{\text{onset}}$
there is a network of interpenetrating domains of the IQHWS and electron liquid.
The presence of the sharp resistive peak in the $R_{xx}(T)|_{\nu=\text{fixed}}$ profiles
as $T$ is scanned
is a result of increased scattering along the domain walls of interpenetrating domains of these two phases.
Similarly, the local maxima in $R_{xx}(B)|_{T=\text{fixed}}$, such as the ones
labeled by $\ast$ symbols  for the $T=55$~mK shown in Fig.\ref{fig:1},
can also be understood as originating from enhanced scattering. We notice that the height of the local resistance 
maxima in the $R_{xx}(B)|_{T=55\text{mK}}$ trace shown in Fig.\ref{fig:1}
is larger when the IQHWS-to-electron liquid boundary is crossed 
at $B=2.71$~T and at $B=3.56$~T and is considerably less across the IQHWS-to-AI boundary 
at $B=2.89$~T and at $B=3.38$~T.

\section{The integer quantum Hall plateau in the $\nu$-$T$ phase space}

\begin{figure}[t]
\includegraphics[width=\columnwidth]{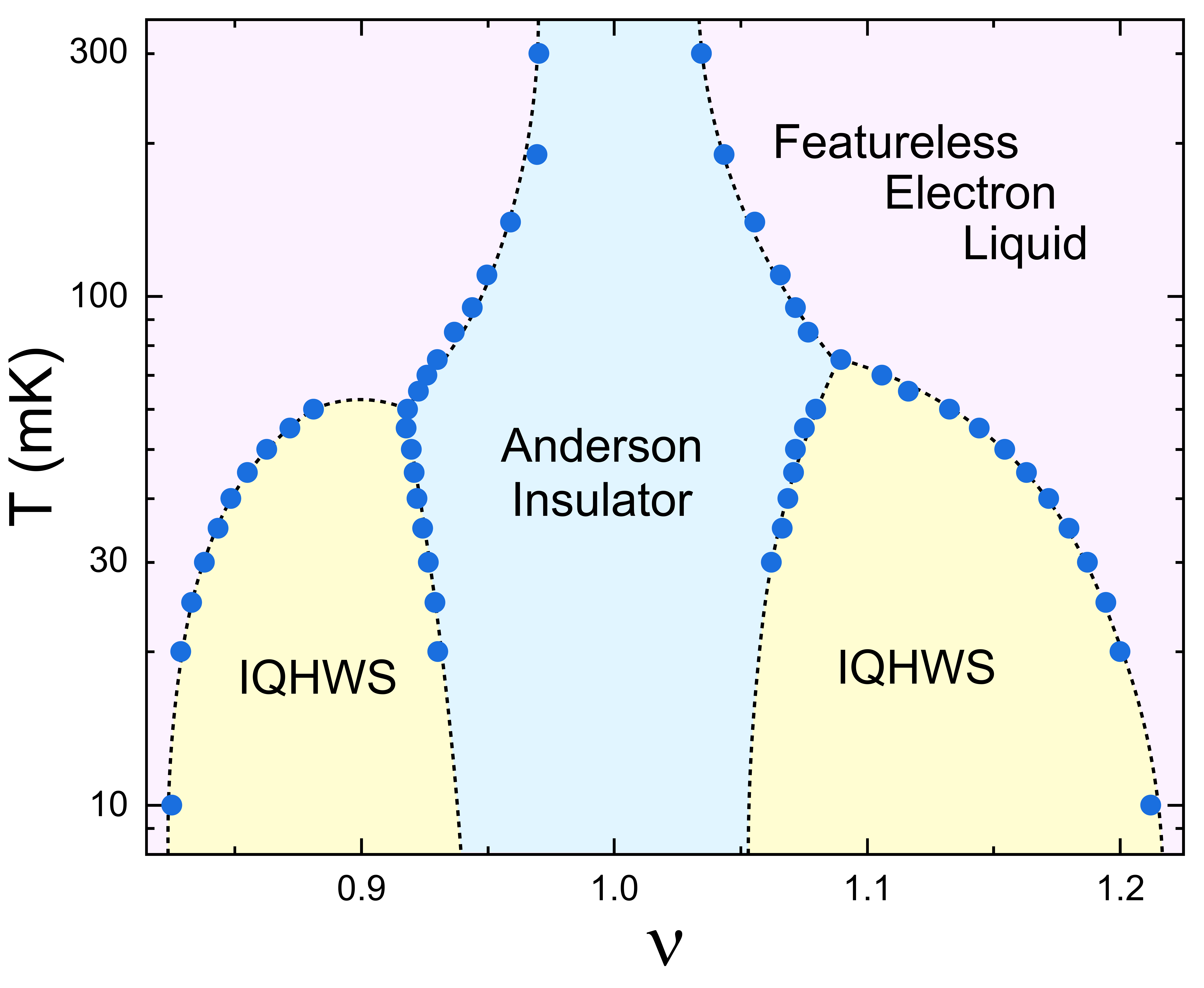} 
\caption{Diagram in $\nu$-$T$ phase space of the localized AI and IQHWS phases
near $\nu=1$. Blue dots mark the filling factors of the local maxima of $R_{xx}(\nu)|_{T=\text{fixed}}$
curves. Dashed lines are guides to the eye.    \label{fig:5} }
\end{figure}

While the data shown in Fig.\ref{fig:3} highlight a sudden change as one crosses the AI-to-IQHWS boundary,
it does not depict this boundary in the $\nu$-$T$ phase space. In order to obtain such a diagram,
we use the local maxima of $R_{xx}(B)|_{T=\text{fixed}}$, such as the ones 
marked by $\ast$ symbols for the $R_{xx}(B)|_{T=55\text{mK}}$ shown in Fig.\ref{fig:1}. 
We associate different localized phases
to the different regions between the local maxima in such traces. As an example, 
the sequence of IQHWS, AI, and IQHWS in the trace of $R_{xx}(B)|_{T=55\text{mK}}$
shown in Fig.\ref{fig:1}
is associated with the following
range of magnetic fields: 2.71~T to 2.89~T, 2.89~T to 3.38~T, and 3.38~T to 3.56~T, respectively.
The blue dots in Fig.\ref{fig:5} represent the boundary
of the localized states obtained this way in the $\nu$-$T$ phase space. The boundaries with the featureless electron
liquid are a measure of the width of the plateau of the $\nu=1$ IQHS. At the highest shown temperature
we have only the AI, which occupies an increasingly wider range of filling factors as the temperature is dropped.
This boundary, however, bifurcates and the width of the integer quantum Hall plateau
increases significantly as the temperature reaches values less than 65~mK, at which point the IQHWS sets in. 
These transport features thus allow us to define a boundary between the AI and the
IQHWS. However, at the lowest temperature reached, transport no longer exhibits any features
to distinguish the AI and IQHWS, and thus for tracing the phase boundary here, other techniques 
will have to be employed. 
Our suggested boundary between the two different localized states at these low temperatures is 
extrapolated from the boundary present at higher temperatures, and it is aided by 
the expectation that this boundary does not shift significantly.

\section{Discussion}

The formation of collective insulators, such as the IQHWS, in the presence of disorder
remains a challenging problem with many unanswered questions. We associated
the reentrant integer quantum Hall effect forming near $\nu=1.1$ and $\nu=0.9$ 
with the IQHWS. However, the absence of
reentrant behavior does not necessarily mean the absence of the IQHWS.
Indeed, with the exception of Ref.\cite{liu-1}, prior transport work on lower mobility samples 
near $\nu=1.1$ and $\nu=0.9$ did not exhibit reentrance, but the IQHWS was found to develop
using techniques other than transport\cite{chen-int,smet-int,suslov-int,ashoori-int}. 
We thus suspect that reentrant transport features studied in this paper develop only in samples with very low
levels of disorder and with the highest mobility. At this time conditions under which IQHWSs
exhibit reentrant transport and thus can be differentiated from the AI in transport measurements
remain unclear. We propose that the morphology of the interpenetrating domains
of the IQHWS and electron liquid, i.e. the domain size of the IQHWS in particular, may play a role.

We found that magnetoresistance patterns near $\nu=2$ are different from those near $\nu=1$.
Indeed, as seen in Fig.\ref{fig:1}, reentrance of the IQHS is absent in our sample near $\nu=2$
down to the lowest temperatures studied. This is puzzling, since IQHWSs were observed near $\nu=2$
in microwave absorption\cite{chen-int}, nuclear magnetic resonance\cite{tiemann-int}, and
surface acoustic wave measurements \cite{suslov-int}
in samples of lower mobility than ours. Other work on the IQHWS
did not examine or report on data near $\nu=2$\cite{liu-1,ashoori-int}. 
We do not have an explanation for the lack of reentrance of the IQHS near $\nu=2$.
One possibility is that the IQHWSs also form there, but in our sample transport signatures associated
with the IQHWS are not present. Such an explanation is consistent with early observation near $\nu=1$,
where microwave spectroscopy found signatures of IQHWSs, but early transport measurements did not
find a reentrant behavior. In an earlier paragraph, we ascribed the presence of enhanced resistance
associated with reentrance as being due to enhanced scattering of interpenetrating domains of an
electron solid and liquid. We think that the lack of reentrance is likely influenced by the
temperature-dependent length scales of these
domains. Another possibility is that the IQHWS does not form near $\nu=2$. The stability of Wigner solids is known
to be affected not just by disorder, but also by sample parameters such as the width of the quantum well
and Landau level mixing. One possibility is thus that in our sample such effects destabilize the IQHWS near $\nu=2$.

It is important to appreciate that the diagram shown in Fig.\ref{fig:5} will depend on the disorder. 
For example, the filling factor ranges for the IQHWS 
and the onset temperatures of the IQHWS will depend on disorder levels.
While it is generally thought that disorder suppresses a collective phase such as the IQHWS,
under special circumstances disorder may stabilize a Wigner solid in a region where the Wigner solid did not
form in the absence of disorder. Such a scenario was realized in samples in which a short
range disorder was embedded into the channel during the molecular beam epitaxy growth
\cite{li-dis,moon-dis} and in which a Wigner solid developed near $\nu \approx 0.6$.
Complex disorder effects are also likely at play in the formation of other collective insulators, such 
as the BPs in high Landau levels\cite{zudov,kevin1}.
In addition, there is ample theoretical \cite{fogler,moessner,fogler96,fogler97,price}
and experimental evidence \cite{liu-1,kevin1,hatke1,hatke2,kate1,kate2}
that the short-range part of the electron-electron interaction tuned
by parameters such as the width of the confining quantum well
will impact the formation of various collective insulators. For example, results of microwave absorption
measurements on samples with significantly lower mobility than ours found structures 
in the flanks of the $\nu=1$ integer plateau that were interpreted as two distinct 
electron solids \cite{hatke1,hatke2}.
We thus expect that such parameters will also impact the stability diagram of the IQHWS.
Lastly, as the disorder is further reduced, new ordered phases may appear in the region of the AI.
However, characterizing the extremely low level of disorder and its spatial distribution
in these samples is not possible with current technologies.

As discussed earlier, the IQHWS and the AI at intermediate temperatures
are separated by a local maximum in the $R_{xx}(B)|_{T=\text{fixed}}$
curves. These local maxima were used to identify the IQHWS and to draw the phase stability diagram 
shown in Fig.\ref{fig:5}. However, at the lowest temperatures we do not detect these local maxima.
This situation is reminiscent of the transport behavior in high Landau levels, where
in a large number of experiments a local maximum was reported separating multi-electron 
BPs from the integer plateau\cite{du,lilly,cooper} or separating different multi-electron BPs \cite{zudov,kevin1}. 
However, there are known exceptions to this rule.  In some samples or for some sample parameters
these local maxima between the BP and the integer plateau are either
not present \cite{b0,b1,b2,b3} or are present at intermediate temperatures but absent at the lowest
temperatures \cite{kevin1,b4,b5,b6}. This latter behavior is similar to the behavior of local maxima separating
the IQHWS and the AI that we observe. Due to the lack of a detailed model for transport, we cannot offer an explanation
for such a behavior. However, we suggest that since the enhanced scattering is due to interpenetrating
domains of competing phases, it is likely that the temperature dependent local resistance maximum
is influenced by the temperature driven change of the domain sizes.

Finally, we note that IQHWSs near $\nu=1$ have spin texture \cite{barrett,bayot,skynmr1,skynmr2,skynmr7,skymicro}. 
Since our measurements are sensitive only to electron localization,
we cannot access any information on magnetic ordering, pertaining for example to skyrmion formation.

In summary, we studied the complex structure of the magnetotransport near the $\nu=1$ IQHS
forming in the newest class of high mobility 2DEGs. Transport in the flanks of the $\nu=1$
exhibits reentrance and the region of reentrance was associated with the IQHWS.
We found that the $R_{xx}(T)_{\nu=\text{fixed}}$ resistance profiles exhibit a sharp peak.
Such sharp peaks in the $R_{xx}(T)_{\nu=\text{fixed}}$ resistance profiles were previously reported
in the BPs, and thus these peaks appear to be present for all isotropic electron solids that form in low-disorder 2DEGs.
The temperature of this sharp peak was assigned to the onset of the IQHWS. Finally, we also 
presented a phase stability diagram in $\nu$-$T$ space of the AI and the IQHWS near the $\nu=1$
integer quantum Hall plateau.

\section*{Acknowledgments}

Measurements at Purdue University were supported by the NSF Grant No. DMR 1904497. Sample growth efforts of L.N.P. and K.W.W. were supported by the NSF MRSEC Grant No. DMR-1420541 and the Gordon and Betty Moore Foundation Grant No. GBMF 4420.



\begin{thebibliography}{l}

\bibitem{laughlin} R.B. Laughlin, Phys. Rev. B \textbf{23}, 5632 (1981).
\bibitem{aoki} H. Aoki and T. Ando, Solid State Communications \textbf{38}, 1079 (1981).
\bibitem{halperin} B. I. Halperin, Phys. Rev. B \textbf{25}, 2185 (1982).
\bibitem{klitz} K. von Klitzing, G. Dorda, and M. Pepper, Phys. Rev. Lett. \textbf{45}, 494 (1980).


\bibitem{fogler} A.A. Koulakov, M.M. Fogler, and B.I. Shklovskii, 
   Phys. Rev. Lett. \textbf{76}, 499 (1996). 
\bibitem{fogler96} M.M. Fogler, A.A. Koulakov, B.I. Shklovskii, 
  Phys. Rev. B \textbf{54}, 1853 (1996). 
\bibitem{moessner} R. Moessner, and J.T. Chalker, 
  Phys. Rev. B \textbf{54}, 5006 (1996). 
\bibitem{fogler97} M.M. Fogler and A.A. Koulakov,
   Phys. Rev. B \textbf{55}, 9326 (1997). 
\bibitem{haldane} F.D.M. Haldane, E.H. Rezayi, and K. Yang,
  Phys. Rev. Lett. \textbf{85}, 5396 (2000). 


\bibitem{chen-int} Y.P. Chen, R.M. Lewis, L.W. Engel, D.C. Tsui, P.D. Ye, L.N. Pfeiffer, and K.W. West, Phys. Rev. Lett. \textbf{91}, 016801 (2003).         
\bibitem{lewis-1} R.M. Lewis, Y. Chen, L.W. Engel, D.C. Tsui, P.D. Ye, L.N. Pfeiffer, and K.W. West, Phys. Rev. Lett. \textbf{93}, 176808 (2004).        
\bibitem{lewis-2} R.M. Lewis, Yong P. Chen, L.W. Engel, D.C. Tsui, P.D. Ye, L.N. Pfeiffer, and K.W. West, Physica E (Amsterdam) \textbf{22}, 104 (2004).    

\bibitem{ws1} H.W. Jiang, R.L. Willett, H.L. Stormer, D.C. Tsui, L.N. Pfeiffer, and K.W. West, Phys. Rev. Lett. 
  \textbf{65}, 633 (1990).
\bibitem{ws2} V.J. Goldman, M. Santos, M. Shayegan, and J.E. Cunningham, Phys. Rev. Lett. \textbf{65}, 2189 (1990).
\bibitem{ws3} chapters by H. A. Fertig and by M. Shayegan in {\it Perspectives in Quantum Hall Effects}, edited by S. Das
Sarma and A. Pinczuk (Wiley, New York, 1997).

\bibitem{lilly} M.P. Lilly, K.B. Cooper, J.P. Eisenstein, L.N. Pfeiffer, and K.W. West, 
  Phys. Rev. Lett. \textbf{82}, 394 (1999) .
\bibitem{du} R.R. Du, D.C. Tsui, H.L. Stormer, L.N. Pfeiffer, K.W. Baldwin, K.W. West, Solid State. Comm. \textbf{109}, 389 (1999).
\bibitem{cooper} K.B. Cooper, M.P. Lilly, J.P. Eisenstein,  L.N. Pfeiffer, and K.W. West, Phys. Rev. B \textbf{60}, R11285 (1999).
\bibitem{eisen02} J.P. Eisenstein, K.B. Cooper, L.N. Pfeiffer and K.W. West, 
  Phys. Rev. Lett. \textbf{88}, 076801 (2002).  
\bibitem{zudov} X. Fu, Q. Shi, M.A. Zudov, G.C. Gardner, J.D. Watson, and M.J. Manfra,
  Phys. Rev. B \textbf{99}, 161402(R) (2019).  
\bibitem{kevin1}  D. Ro, N. Deng, J.D. Watson, M.J. Manfra, L.N. Pfeiffer, K.W. West, and G.A. Cs\'athy,
Phys. Rev. B \textbf{99}, 201111(R) (2019).

\bibitem{dean}S. Chen, R. Ribeiro-Palau, K. Yang, K. Watanabe, T. Taniguchi, J. Hone, M. O. Goerbig, and C. R. Dean, Phys. Rev. Lett. \textbf{122}, 026802 (2019).
\bibitem{young} H. Zhou, H. Polshyn, T. Taniguchi, K. Watanabe, and A.F. Young,
Nat. Physics \textbf{16}, 154 (2020).

\bibitem{smet-int} D. Zhang, X. Huang, W. Dietsche, K. von Klitzing, and J.H. Smet, Phys. Rev. Lett. \textbf{113}, 076804 (2014).       
\bibitem{tiemann-int} L. Tiemann, T.D. Rhone, N. Shibata, and K. Muraki, Nat. Phys. \textbf{10}, 648 (2014).                      
\bibitem{suslov-int} I.L. Drichko, I.Yu. Smirnov, A.V. Suslov, L.N. Pfeiffer, K.W. West, and Y.M. Galperin, Phys. Rev. B \textbf{92}, 205313 (2015).             
\bibitem{ashoori-int} J. Jang, B.M. Hunt, L.N. Pfeiffer, K.W. West, and R.C. Ashoori, Nature Phys. \textbf{13}, 340 (2017).                 
\bibitem{liu-1} Y. Liu, C.G. Pappas, M. Shayegan, L.N. Pfeiffer, K.W. West, and K.W. Baldwin, Phys. Rev. Lett. \textbf{109}, 036801 (2012).  


\bibitem{chung} Y.J. Chung, K.A. Villegas Rosales, K.W. Baldwin, P.T. Madathil, K.W. West, M. Shayegan, and L.N. Pfeiffer, Nat. Mater. \textbf{20}, 632 (2021).

\bibitem{deng-1} N. Deng, A. Kumar, M.J. Manfra, L.N. Pfeiffer, K.W. West, and G.A. Cs\'athy,
  Phys. Rev. Lett. \textbf{108}, 086803 (2012).  
\bibitem{deng-2} N. Deng, J.D. Watson, L.P. Rokhinson, M.J. Manfra, and G.A. Cs\'athy, 
  Phys. Rev. B \textbf{86}, 201301(R) (2012).   
\bibitem{chick} W.E. Chickering, PhD Thesis, California Institute of Technology (2015).
\bibitem{kevin2} D. Ro, S.A. Myers, N. Deng, J.D. Watson, M.J. Manfra, L.N. Pfeiffer, K.W. West, and G.A. Cs\'athy, Phys. Rev. B \textbf{102}, 115303 (2020).  

\bibitem{setup} N. Samkharadze, A. Kumar, M.J. Manfra, L.N. Pfeiffer, K.W. West, and G.A. Cs\'athy, Rev. Sci. Instrum. \textbf {82} 053902 (2011).

\bibitem{li-dis} W. Li, D.R. Luhman, D.C. Tsui, L.N. Pfeiffer, and K.W. West, 
  Phys. Rev. Lett. \textbf{105}, 076803 (2010).    
\bibitem{moon-dis} B.-H. Moon, L.W. Engel, D.C. Tsui, L.N. Pfeiffer, and K.W. West, 
  Phys. Rev. B \textbf{92}, 035121 (2015).    
\bibitem{price} R. Price, X. Zhu, S. Das Sarma, and P.M. Platzman, Phys. Rev. B \textbf{51}, 2017 (1995).
\bibitem{hatke1} A.T. Hatke, Y. Liu, B.A. Magill, B.H. Moon, L.W. Engel, M. Shayegan, L.N. Pfeiffer, K.W. West, and
K.W. Baldwin, Nature Commun. \textbf{5}, 4154 (2014).
\bibitem{hatke2} A.T. Hatke, Y. Liu, L.W. Engel, L.N. Pfeiffer, K.W. West, K.W. Baldwin, 
and M. Shayegan, Phys. Rev. B \textbf{98}, 195309 (2018).
\bibitem{kate1} K.A. Schreiber, N. Samkharadze, G.C. Gardner, Y. Lyanda-Geller, M.J. Manfra, L.N. Pfeiffer,
K.W. West, and G.A. Cs\'athy, Nature Commun. \textbf{9}, 2400 (2018).
\bibitem{kate2}  K.A. Schreiber and G.A. Cs\'athy, Annual Rev. Cond. Mat. Phys. \textbf{11}, 17 (2020).

\bibitem{b0} K.B. Cooper, M.P. Lilly, J.P. Eisenstein, L.N. Pfeiffer, and K.W. West,
   Physical Review B \textbf{65}, 241313(R) (2002).
\bibitem{b1} Y. Liu, D. Kamburov, M. Shayegan, L.N. Pfeiffer, K.W. West, and K.W. Baldwin,
  Physical Review B \textbf{87}, 075314 (2013).
\bibitem{b2} G. Gamez and K. Muraki,
  Physical Review B \textbf{88}, 075308 (2013).
\bibitem{b3} B. Friess, Y. Peng, B. Rosenow, F. von Oppen, V. Umansky, K. von Klitzing, and J.H. Smet,
  Nature Phys. \textbf{13},1124  (2017).
  
\bibitem{b4} J.P. Eisenstein, M.P. Lilly, K.B. Cooper, L.N. Pfeiffer, K.W. West,
  Physica E \textbf{6}  29 (2000).  
\bibitem{b5} X. Wang, H. Fu, L. Du, X. Liu, P. Wang, L.N. Pfeiffer, K.W. West, R.R. Du, and X. Lin,
  Phys. Rev. B \textbf{91}, 115301 (2015).        
\bibitem{b6} K. Bennaceur, C. Lupien, B. Reulet, G. Gervais, L.N. Pfeiffer, and K.W. West,
  Phys. Rev. Lett. \textbf{120}, 136801 (2018).      

\bibitem{barrett} S.E. Barrett, G. Dabbagh, L.N. Pfeiffer, K.W. West, and R. Tycko, Phys. Rev. Lett. \textbf{74}, 5112 (1995).
\bibitem{bayot} V. Bayot, E. Grivei, S. Melinte, M.B. Santos, and M. Shayegan, Phys. Rev. Lett. \textbf{76}, 4584 (1996).
\bibitem{skynmr1} W. Desrat, D. K. Maude, M. Potemski, J.C. Portal, Z.R. Wasilewski, and G. Hill, Phys. Rev. Lett. \textbf{88}, 256807 (2002).
\bibitem{skynmr2} G. Gervais, H.L. Stormer, D.C. Tsui, P.L. Kuhns, W.G. Moulton, A.P. Reyes, L.N. Pfeiffer, K.W. Baldwin, and K.W. West, Phys. Rev. Lett. \textbf{94}, 196803 (2005).
\bibitem{skymicro} H. Zhu, G. Sambandamurthy, Y.P. Chen, P. Jiang, L.W. Engel, D.C. Tsui, L.N. Pfeiffer, and K.W. West, Phys. Rev. Lett. \textbf{104}, 226801 (2010).
\bibitem{skynmr7} W. Desrat, B.A. Piot, S. Kr\"{a}mer, D.K. Maude, Z.R. Wasilewski, M. Henini, and R. Airey, Phys. Rev. B \textbf{88}, 241306(R) (2013).

\end{thebibliography}
\end{document}